\documentclass{aa}

\usepackage{graphicx}
\usepackage{txfonts}

\begin{document}

\title{Photometric variability of the Herbig Ae star
HD~37806\thanks{Based on data from the MOST
satellite, a Canadian Space Agency mission jointly
operated by Dynacon Inc., the University of Toronto Institute for
Aerospace Studies and the University of
British Columbia, with the assistance of the University
of Vienna, and on data from the All Sky Automated Survey (ASAS)
conducted by the Warsaw University Observatory, Warsaw, Poland
at the Las Campanas Observatory, Chile.
}}

\titlerunning{Photometric variability of HD~37806}
\authorrunning {S. M. Rucinski et al.}

\author{S. M. Rucinski\inst{1} \and
K. Zwintz\inst{2} \and
M. Hareter\inst{2} \and
G. Pojma\'{n}ski\inst{3} \and
R. Kuschnig\inst{2} \and
J. M. Matthews\inst{4} \and
D. B. Guenther\inst{5} \and
A. F. J. Moffat\inst{6} \and
D. Sasselov\inst{7} \and
W. W. Weiss\inst{2}
}

\institute{
Department of Astronomy and Astrophysics,
    University of Toronto, 50 St.\ George St., Toronto,
    Ontario, M5S~3H4, Canada\\
    \email rucinski@astro.utoronto.ca                                   \and
Institut f\"ur Astronomie, Universit\"at Wien,
    T\"urkenschanzstrasse 17, A-1180 Vienna, Austria \\
\email konstanze.zwintz@univie.ac.at, markus.hareter@univie.ac.at,
rainer.kuschnig@univie.ac.at, werner.weiss@univie.ac.at                 \and
Warsaw University, Astronomical Observatory,
     Al.\ Ujazdowskie 4, 00--478 Warszawa, Poland\\
    \email gp@astrouw.edu.pl                                            \and
Department of Physics and Astronomy, University of British Columbia,
    6224 Agricultural Road, Vancouver, BC V6T 1Z1, Canada \\
    \email matthews@astro.ubc.ca                                        \and
Department of Astronomy and Physics, St. Mary's University, Halifax,
    NS B3H 3C3, Canada \\
    \email guenther@ap.smu.ca                                             \and
D\'epartment de physique, Universit\'e de Montr\'eal, C.P. 6128,
    Succ. Centre-Ville, Montr\'eal, QC H3C 3J7, Canada \\
    \email moffat@astro.umontreal.ca                                      \and
Harvard-Smithsonian Center for Astrophysics, 60 Garden Street,
    Cambridge, MA 02138, USA \\
    \email sasselov@cfa.harvard.edu
}

\date{Received / Accepted }

\abstract         
{The more massive counterparts of T~Tauri stars,
Herbig Ae/Be stars, are known to vary in a complex way
with no variability mechanism clearly identified.}
{We attempt to characterize the optical variability of HD~37806 (MWC~120)
on time scales ranging between minutes and several years.}
{A continuous, one-minute resolution, 21 day-long sequence of
MOST (Microvariability \& Oscillations of STars) satellite observations
has been analyzed using wavelet, scalegram and dispersion
analysis tools. The MOST data have been augmented by sparse
observations over 9 seasons from ASAS (All Sky Automated Survey),
by previously non-analyzed ESO (European Southern Observatory) data
partly covering 3 seasons and by archival measurements dating
back half a century ago.}
{Mutually superimposed flares or accretion instabilities grow
in size from about 0.0003 of the mean flux on a time scale of
minutes to a peak-to-peak range of $<$~0.05 on a time scale of a few
years. The resulting variability has properties of
stochastic ``red'' noise, whose self-similar characteristics are very
similar to those observed in cataclysmic binary stars,
but with much longer characteristic time scales of hours to days
(rather than minutes) and with amplitudes which appear to
cease growing in size on time scales of tens of years.
In addition to chaotic brightness variations combined with
stochastic noise, the MOST data show a weakly defined
cyclic signal with a period of about 1.5 days, which may
correspond to the rotation of the star.}
{}


\maketitle

\titlerunning{Photometric variability of HD~37806}
\authorrunning{S. M. Rucinski et al.}


\section{Introduction}
\label{intro}

HD~37806 (SAO~132452, RA(2000) = 05:41:02.3, DEC(2000) = $-$02:43:01,
$V \simeq 7.95$) is an early star in the field of Sigma Ori.
It is frequently referred to as MWC~120, the designation
in the Merrill \& Burwell Catalog (\cite{mb1933})
of early type stars with emission lines, which it received
following the discovery of Merrill et al. (\cite{mhb1932}).
HD~37806 is a Herbig Ae/Be-type object, i.e.\ a young star
which is more massive (typically $1.5 - 5 \, M_\odot$)
than solar and sub-solar mass T~Tauri stars.
The review of Waters \& Waelkens (\cite{WW1998})
summarizes the properties of Herbig Ae/Be-type stars;
a catalogue of Herbig Ae/Be-type objects was
published by Th\'{e} et al.\ \cite{The1994}).
The MK spectral type of HD~37806 is variously given
as A2Vpe or B9Ve+sh (Swings \& Struve \cite{SS1943},
Guetter \cite{guet1981}, Th\'{e} et al.\ \cite{The1994}) although
sometimes it is still quoted as A0.
HD~37806 was one of the first ``radio'' stars identified
(Feldman et al.\ \cite{feld1973}).

The age and cluster membership of HD~37806 are not well known,
although the star does appear very young.
While it is only 35 arcmin from Sigma Orionis, it was not included
among the members of the young cluster carrying the name of this
star in the study by Sherry et al.\ (\cite{sher2004,sher2008}).
The distance estimated for the cluster, $420 \pm 30$ pc,
is consistent with the Hipparcos new-reduction parallax
$2.05 \pm 0.91$ mas (van Leeuwen \cite{new-hip}).
The star almost certainly belongs to the Orion OB1b association
(of which the Sigma Ori cluster may be a part), whose age is estimated at
2 -- 5 Myr (Brown et al.\ \cite{brown1994}) and for which the Hipparcos
group distance is $350_{-90}^{+120}$ pc.
In Brown et al.\ (\cite{brown1994}), the distance to HD~37806
was estimated at about 300 pc, assuming
$V = 7.94$, $M_V = +0.2$, and a moderately large $A_V = 0.35$;
for the Sigma Ori cluster, Sherry et al.\ (\cite{sher2004,sher2008})
estimate $A_V$ at about 0.1 to 0.2. The distance,
the possible membership of HD~37806 to the Sigma Ori
cluster and the amount of interstellar
absorption all appear to be very uncertain at this point.

The main thrust of this paper is limited to
utilizing the continuous, 21 day long observations by the MOST
satellite for characterization of the optical variability
of HD~37806. As auxiliary data, we use lower
accuracy ESO and ASAS observations, which extend
the time coverage to several years, and archival observations
extending the time baseline to half a century.
The data analysis tools are similar to those used for
continuous MOST satellite observations
of the T~Tauri type star TW~Hya (Rucinski et al.\ \cite{ruc2008}).

The literature on diversified studies of HD~37806 in various
spectral regions is extensive, where the star is usually identified
by its more popular alias MWC~120. Among the recent studies,
Mottram et al.\ (\cite{mott2007}) pointed out
that Herbig Ae stars (A0--A2), such as HD~37806,
appear to be more similar in terms of the accretion
processes producing the emission spectra
to T~Tauri stars than to earlier (B0--B3) Herbig Be  stars.
The star has also been the subject of extensive spectroscopic,
polarimetric and spectro-polarimetric investigations
(Vink et al. \cite{Vink2005}). Of relevance to variability
studies of HD~37806 is the finding
by Harrington \& Kuhn (\cite{HK2009})
that the H$\alpha$ line shows strong morphological changes
on a time scale of 3 years.

Very recently, using a novel method of spectro-astrometric signatures,
Wheelwright et al.\ (\cite{Wh2010})
found that HD~37806 is a binary system, similarly to the majority
of surveyed 45 Herbig Ae/Be stars. This is a
very important discovery which may provide the key explanation of
the variability characteristics which we see in our MOST data;
we return to this subject in Sections~\ref{disc} and \ref{summ}.

\section{Optical variability: Previous photometric results}
\label{prev}

\subsection{Archival data}
\label{arch}

HD~37806 was observed in the Johnson $V$ and Str\"omgren $y$
(transformed to $V$) bands in 1959 -- 1962 and 1968 by
Hardie et al.\ (\cite{Hard1964}) and Warren \& Hesser (\cite{WH1977}).
Additionally, de Winter et al.\ (\cite{deW2001}) observed
the star in 1986 and 1992.
Although all these were few (2 and 5) scattered  observations,
they give an interesting constraint on the
long term variability of the star (Section~\ref{disc}).

\subsection{Hipparcos observations}

Hipparcos satellite observations of 44
Herbig Ae/Be stars, obtained over 37 months
of the mission in the $Hp$ band,
were analyzed by van den Ancker (\cite{vdA1998}).
Typically a hundred time-dispersed $H_P$ magnitude
determinations per star were obtained with accuracy of
0.008 -- 0.015 mag. In this study,
HD~37806 was observed to vary by about 0.07 mag.
However, this study was obviously unable to define
detailed variability characteristics of the star.

We looked at the Hipparcos data again.
The data consist of 135 $H_P$ observations and
209 $V_T$ and $B_T$ observations of the Tycho project,
the latter with rather large uncertainties of typically
0.06 -- 0.12 mag. All observations
tend to be strongly bunched in time as a result of the
sky scanning pattern of the satellite and appear in only
a few groups over the mission duration.
The strong bunching would not be a detrimental obstacle for
use in the long time-scale variability considerations,
as in Section~\ref{disc},
if the main-project data were in the $V$ rather than $H_P$
system.

In the end, we used only the Tycho-2
(H{\o}g et al. \cite{Tycho2}) well-calibrated, single
data point corresponding to the average $V=7.934 \pm 0.016$ (note that
there exists no Tycho-2 epoch photometry). It has been added to the
remaining archival observations discussed later on
(Section~\ref{disc}) in an attempt to relate variability
seen in by us to its possible long-term continuation.

\subsection{ESO observations}
\label{eso}

Four-color Str\"{o}mgren $uvby$ observations of HD~37806 were
conducted at the European Southern Observatory (ESO) in
1985 -- 1986 by several observers (Manfroid et al.\ \cite{eso});
the observations do not appear to have been analyzed so far.
Altogether 55 observations in the $y$ ($= V$) band were
obtained, with 12 during the first season of
January -- February 1985, 41 observations during the
November 1985 -- March 1986 season and with
2 observations in September 1986.
The typical interval between observations was 2 days and
there were no multiple observations taken on the same night.
The observations were done differentially and were
relatively accurate with typical measurement errors of
0.007 mag.

\begin{figure}
\begin{center}
\includegraphics[width=80mm]{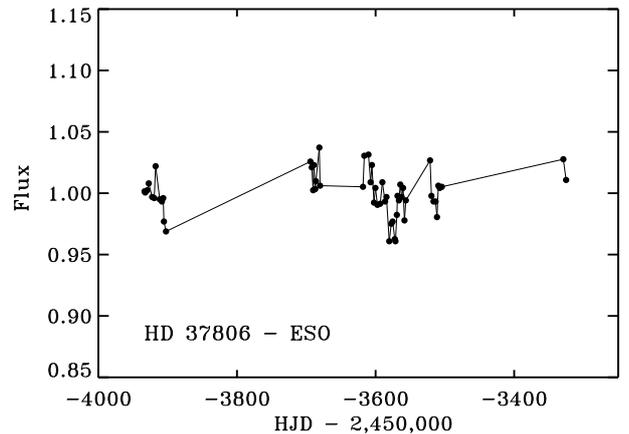}
\caption{The ESO light curve of HD~37806 in 3 seasons of 1985 -- 1986,
expressed in units of the
mean flux. The central grouping of points corresponds to one
Chilean observing season; the two other seasons were poorly covered.
The HJD offset used in this figure is the same as for the
remaining plots in this paper, hence the negative values of time.
}
\label{ESOlc}
\end{center}
\end{figure}

\begin{figure}
\begin{center}
\includegraphics[width=80mm]{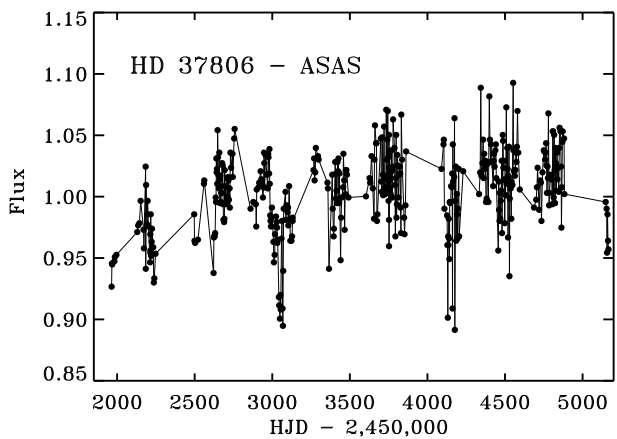}
\caption{The ASAS light curve of HD~37806 expressed in units of the
mean flux. The groupings of points correspond to
individual observing seasons. Note the different horizontal
scale compared with Figure~\ref{ESOlc}.
}
\label{ASASlc}
\end{center}
\end{figure}

\begin{figure*}
\begin{center}
\includegraphics[width=120mm]{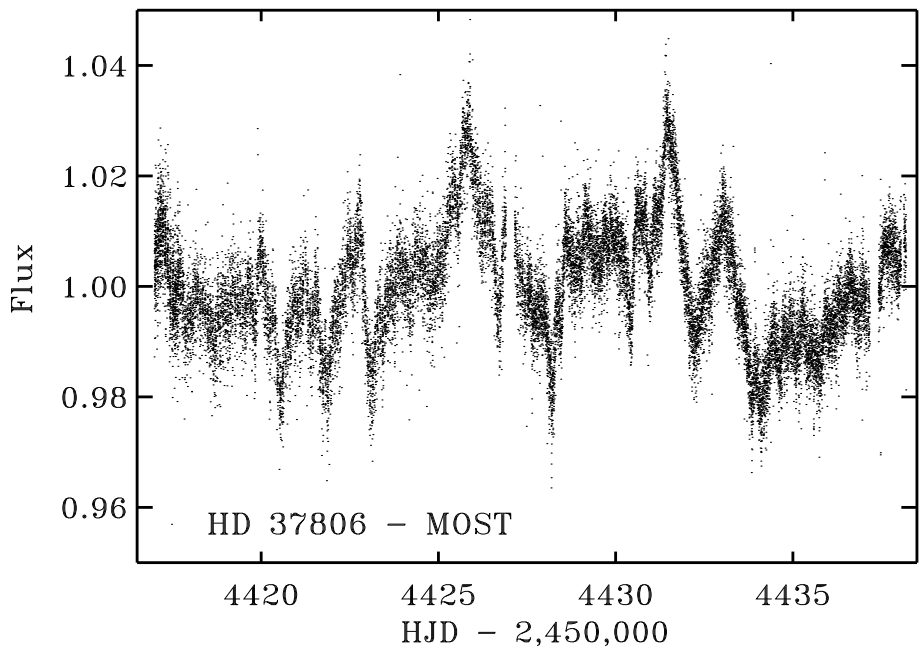}
\caption{The MOST light curve of HD~37806, covering 21 days,
expressed in units of the mean flux, with
one-minute sampling. The first half of the run
experienced gaps of about 30\% of each 103
minute satellite orbit due to a high background level.
The data set has been used at the full one-minute
resolution as well as at the uniform, MOST-orbit time sampling
necessary for the wavelet analysis (see Section~\ref{wavelet}.)}
\label{MOSTlc}
\end{center}
\end{figure*}

The ESO observations of HD~37806 are shown in Figure~\ref{ESOlc}.
The observations do not indicate any regularity in its variability, but
rather random noise, perhaps with increasing amplitudes
for time scales ranging from single days to hundreds of days.
We discuss the optical flux scatter in Section~\ref{noise},
where we interpret the variations in terms of stochastic noise
characterized by $\sigma_\star ^2 = \sigma_{obs}^2 - \sigma_R^2$,
where the observational (instrumental) noise is described by the
contribution $\sigma_R = 0.007$.

\subsection{ASAS observations}
\label{asas}

ASAS, the All Sky Automated Survey (Pojma\'{n}ski
\cite{Poj1997,Poj2002,Poj2004,Poj2005,BP2006}\footnote{See:
http://www.astrouw.edu.pl/asas/ }), is a long
term project dedicated to detection and monitoring of
variability of bright stars using small telescopes.
It has been run for a decade by Warsaw University at Las Campanas
Observatory in Chile using 7~cm telescopes providing
photometry in the 8 -- 13 magnitude range.
About three fourths of the sky
($-90^\circ < \delta < +28^\circ$) is covered on a regular basis.

The ASAS observations of HD~37806, 424 in number, were obtained
in the $V$ filter over 9 seasons, from February 2001
to December 2009. They are shown in Figure~\ref{ASASlc}.
The typical random errors are somewhat difficult to
characterize and were typically at the level of 0.015 -- 0.02 mag.
The errors include a major component resulting
from reductions of individual frames to the $V$ band, which
are done relative to all available Tycho-2 catalogue
stars within the field of view of the telescope. While this
approach assures the photometrically differential nature of
the data (with possible systematic errors within the frame),
it may generate -- due to environmental factors such as
clouds crossing the large field of view,
extinction anomalies -- random errors when individual
frames are compared, as in variability studies. Also,
the inexpensive (engineering grade CCD) detector
had many bad and warm pixels which were observed to
vary their response in time and were sometimes not
immediately recognized in automatic
reductions. These issues, coupled with the rather small variability
amplitude for HD~37806 forced re-reductions of the data for
this star which are slightly different from those available
from the automatic on-line ASAS data system.

The observed brightness variations of HD~37806
are dominated by random stellar flux fluctuations, enlarged
by the observational noise. In the analysis
of the brightness fluctuations in Section~\ref{noise}
we assume observational random noise of $\sigma_R = 0.017$.
The two well defined brightness decreases
(at time 3050 and 4100 in Figure~\ref{ASASlc})
appear genuine as we could not identify any instrumental
effects to produce them; they may be manifestations of
sudden obscuration brightness drops observed for the UX~Ori-type
subclass of Herbig Ae/Be stars (Waters \& Waelkens \cite{WW1998}).

\section{MOST satellite observations}
\label{most}

\subsection{The data}

Originally planned for two years, the MOST satellite has been
observing variability of stars highly successfully since 2003.
The optical system of the satellite consists
of a Rumak-Maksutov f/6 15~cm reflecting telescope.
The custom broad-band filter covers the spectral range of
380 -- 700~nm with effective wavelength located close
to Johnson's $V$ band. The pre-launch characteristics of the
mission are described by Walker et al.\ (\cite{WM2003})
and the initial post-launch performance by Matthews et al.\
(\cite{M2004}).

HD~37806 was observed by MOST in Guide Star mode for 21 days,
during observations of Sigma Orionis and its surrounding
stars in November 2007 (Townsend et al.; in preparation).
The data were reduced with other stars of this
campaign using the MOST Guide Star Photometry reduction method
(Hareter et al.\ \cite{haret2008}) and expressed as added counts
per one minute interval. In this
paper we utilize only the flux values corresponding to
counts normalized to their mean value. The mean count rates
are not relevant for error estimation
because of the highly non-Poissonian
noise characteristics resulting from addition of frequent
CCD readouts (typically every second) of the Guide Star mode
observations.

Continuous, uninterrupted, one-minute interval observations
were done during the second half of the run. Thus, a full
analysis at this temporal resolution could be done
over a shorter interval of 9.5 days only (Section~\ref{noise}).
During the first half
of the run, the observations had stray-light elimination
gaps of typically 29 minutes per 103 minute MOST orbit.
In this paper, we used both the whole dataset sampled
at one minute as well as ``normal points'' obtained by
median averaging of naturally formed groups of observations
at MOST-orbit intervals of 0.07 d.
The orbital normal points were used for the wavelet analysis
(Section~\ref{wavelet}) which requires data sampled
at strictly equidistant points in time.

Well-defined fluctuations of a few percent from the mean flux are
visible in the MOST light curve (Figure~\ref{MOSTlc}).
The fluctuations appear to be reminiscent of
flares or irregular brightening events observed in cataclysmic
and nova-like variables (CV) and which are explained as stochastic
variability resulting from superposition of overlapping
accretion instability events in the disk surrounding the star.

\begin{figure}
\begin{center}
\includegraphics[width=80mm]{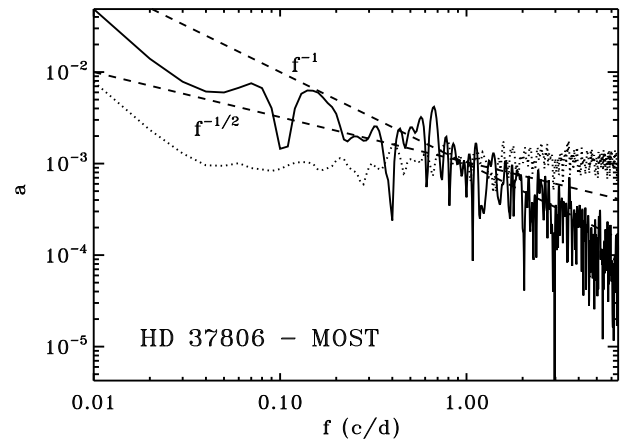}
\caption{The amplitude spectrum of the MOST variability data for
HD~37806 is shown by the continuous line. The dotted line gives
the Fourier amplitude errors estimated by a bootstrap re-sampling
experiment.
The broken lines, shown to guide the eye, give the slopes of
$a \propto 1/f$ and $a \propto 1/\sqrt{f}$; the latter corresponds
to ``flicker noise'' (Press \cite{Press1978}).
}
\label{MOSTfreq}
\end{center}
\end{figure}

Fourier analysis of the MOST data was done for all available data
sampled at one minute intervals by fitting cosine-sine
pairs with frequencies incremented in small steps to cover the
whole accessible frequency range. This corresponded to
a range extending from the lowest accessible
0.05 cycles per day (c/d) to several hundred cycles per day.
The amplitude spectrum, shown in
Figure~\ref{MOSTfreq} in logarithmic units,
is dominated by low-frequency variability at frequencies
below a few cycles per day. Usually, due to stray-light
effects, passages through the South Atlantic Anomaly, etc.\
the MOST orbit frequency of 14.2 c/d, along with a
frequency of 1 c/d appear in such spectra but both are absent in the
reduced material, attesting to good processing of the raw data.
The low frequency content has led us to use in the
wavelet analysis (Section~\ref{wavelet})
the data averaged into points sampled at the MOST-orbit
frequency, i.e.\ below 7 c/d.

In addition to the Fourier component amplitudes,
Figure~\ref{MOSTfreq} gives the mean standard errors for the
amplitudes,
estimated using a bootstrap experiment of multiply re-sampled data.
The errors start to become significant for frequencies higher
than a few c/d; their value was used to estimate the average
mean random error for the MOST data at $\sigma_R = 0.0028$
of the mean flux.

No coherent, periodic signal is visible
in the amplitude spectrum except
for a somewhat complex feature at about 0.65 c/d.
The overall spectrum appears to be that of red stochastic noise,
possibly even ``redder'' (with stronger low frequencies)
than the common ``flicker noise'' (amplitudes
scaling as $a \propto 1/\sqrt{f}$; Press \cite{Press1978});
the slope of the amplitude spectrum
may in fact show a change around the discrete spectral
feature at $\simeq 0.65$ c/d or the period about 1.5 days
(see Figure~\ref{MOSTfreq}).

\begin{figure}
\begin{center}
\includegraphics[width=80mm]{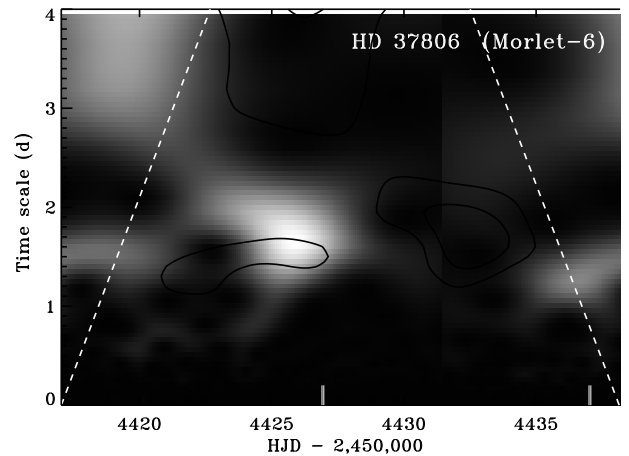}
\caption{The Morlet-6 wavelet (sine-cosine pair extending
over 6 cycles and confined by a Gaussian)
applied to the HD~37806 data sampled
at the MOST orbital period of 0.07046 d.
The light gray areas correspond to  stronger power.
Edge effects influence the wavelet transform beyond the broken lines.
Note that the phase information for this complex-function transform has been
disregarded by using the power of the transform. The missing orbital
data were interpolated at two points marked by
tick marks along the lower horizontal axis, but no detrimental
effects appear to be present there.
}
\label{Morlet-6}
\end{center}
\end{figure}

\subsection{Wavelet transforms}
\label{wavelet}

The data for the wavelet transform must be arranged
into a strictly equidistant time sequence to accommodate progressive
extension of the temporal
scale by integer multiples of the basic sampling interval.
For MOST data, the most
obvious such sampling interval is the duration of one satellite
orbit around the Earth. Indeed, the majority of
instrumental and stray-light problems usually manifest themselves
with a period of 0.07046 d.
The median flux values for HD~37806 from individual satellite orbits
were obtained from typically 70 to 100 data points per median
in the first half of the run
(when orbital gaps were present) and up to 102 points (minus
outliers and glitches)
per median in the second half (the last 9.5 days) of the run.
The resulting data were additionally spline interpolated -- with
very slight shifts of less than a minute --
into a strictly equidistant time grid with steps of
0.07046 d. Two gaps in the mean data were present on $HJD$ -- 2,450,000
days 4427 and 4437; they were filled by spline
interpolated points to obtain a uniform sequence of 301 data points.

Wavelet analysis is normally done with time-scale
incremented as a power of 2, i.e.\ in 2, 4, 8, 16...
data sampling intervals (here 0.07046 d each). The same sequence
was used here, but -- for convenience of the
subsequent transform gray-scale displays -- the scales were later
interpolated into a linear sequence extending from 0.14 to 4 days;
for still longer time scales, edge effects start becoming important
for the MOST data.

Various types of wavelet functions are in widespread use in data
processing (Torrence \& Compo \cite{TC1998}).
Selection of a particular type
depends on the type of variability. The simplest and most
often used are the Morlet wavelet of a sine-cosine pair
confined in a Gaussian wave ``packet'', which is
useful for detection of short-lasting or evolving periodic
events, and the Derivative of a Gaussian (DOG) wavelet,
in its lowest-order version (the second derivative, DOG-2),
which is sometimes called the Mexican Hat wavelet and which
is useful for characterization of chaotic, localized
events. We used both because
variability of HD~37806 appears to be dominated by random,
chaotically occurring flares, yet the lesson of
TW~Hya (Rucinski et al.\ \cite{ruc2008}) taught
us that such stars may show hidden, evolving periodicities.
For the Morlet wavelet, we used the 6-th order version because
it was found in the analysis of TW~Hya that the more commonly used Morlet-5
did not reproduce the Fourier frequencies sufficiently well (and this
has been confirmed again here); qualitatively,
the results for both Morlet transforms are very similar.

The Morlet-6 transform (Figure~\ref{Morlet-6}) confirms the presence
of a quasi-periodic signal with a period of about 1.5 d, which
was already noted in the Fourier analysis (Figure~\ref{MOSTfreq}). In the
wavelet transform, the 1.5 d signature appeared twice during
the three weeks of the observations, with wavelet amplitudes
0.027 and 0.040 of the mean flux, which can be compared
with the amplitude of a bit less than 0.003 in the whole, ``diluted-in-time''
Fourier spectrum. In Section~\ref{disc} we interpret the 1.5 day period
as related to the rotation of the star with
$V \sin i = 120 \pm 30$ km/s (Boehm \& Catala \cite{BC1995}).

\begin{figure}
\begin{center}
\includegraphics[width=80mm]{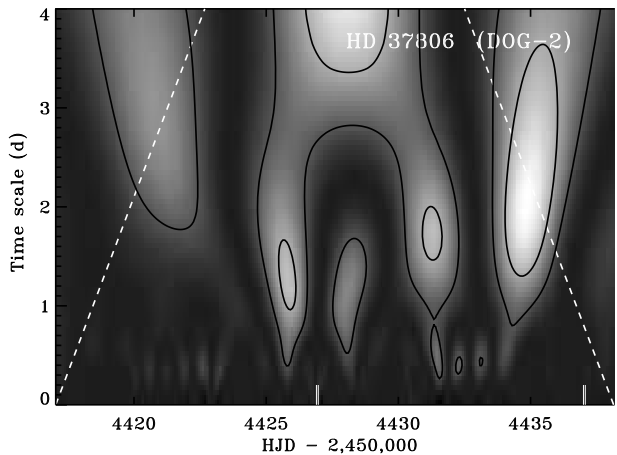}
\caption{The same as in Figure~\ref{Morlet-6}, but
for the DOG-2 wavelet (the negative second derivative of
a Gaussian).  Note that the shortest coherent noise
structures seem to appear at time scales of about 0.25 d.
}
\label{DOG-2}
\end{center}
\end{figure}

\begin{figure}
\begin{center}
\includegraphics[width=85mm]{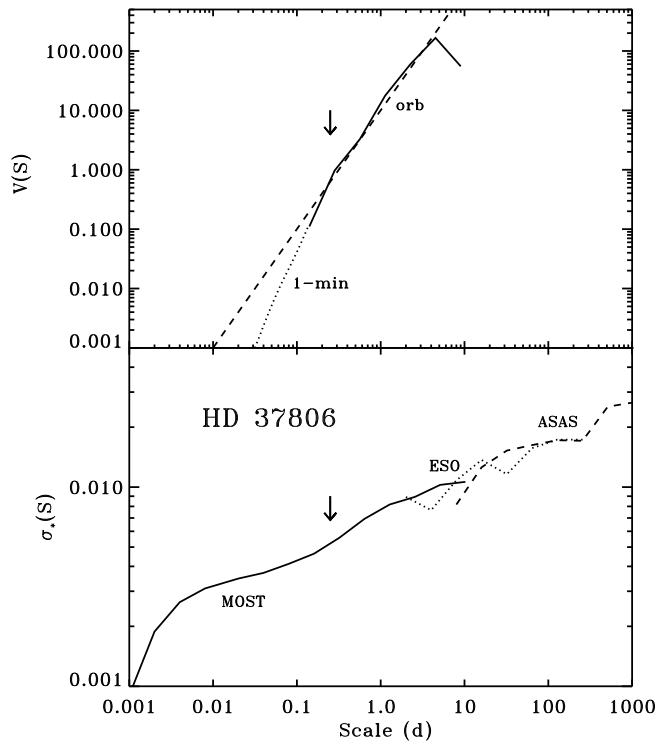}
\caption{{\it Upper panel:}
The scalegram for the 21 days of the MOST data for
HD~37806 based on the DOG-2 wavelet (continuous line) for
the data were sampled
at the MOST orbital period of 0.07 d. The horizontal
axis gives the time scale in days. Because only slope
values and their changes are important here, the vertical scale
has an arbitrary zero point. The broken line has slope
3 in log--log units. The dotted line gives the scalegram
for the shortest time scales,  based on 9.5 days of data
sampled at one-minute intervals (see Figure~\ref{DOG-2short});
it has been shifted vertically to match the results for the
coarsely sampled data. The arrow points at
time scales where visual inspection suggests appearance of
coherent structures in the DOG-2 wavelet transform.
{\it Lower panel:} The continuous line gives the
median standard deviation $\sigma_\star$
for the full set of the MOST data (sampled at one minute),
calculated for progressively longer
time scales and corrected for the observational random
error of $\sigma_R=0.0028$, as described in the text.
The dotted and broken lines give the ESO and the ASAS
median dispersions at a given scale,
corrected for observational errors $\sigma_R$
of 0.007 and 0.017, respectively.
}
\label{scale_disp}
\end{center}
\end{figure}

The DOG-2 wavelet transform (Figure~\ref{DOG-2}),
which is an excellent tool for studies of chaotic noise, shows
a different picture from the Morlet-6 transform: We see
individual brightening events which become larger for longer time scales.
Some events die out quickly or may be just
observational noise effects, but some grow in
brightness for longer time scales. The growing strength
with  temporal scale is a characteristic feature for a self-similar
noise process. This is discussed in the next section.

\subsection{Brightness variability as stochastic noise}
\label{noise}

An underlying stochastic process is considered to be
self-similar with similarity exponent $\alpha$
when it behaves as: $X(\lambda t) = \lambda^\alpha X(t)$.
Scargle et al.\ (\cite{scarg1993}) considered
noise characteristics of accretion sources
and introduced the concept of a ``scalegram'' function
to measure the self-similarity properties of such processes.
The same approach with an extensive discussion and
utilization of wavelets for cataclysmic variable
data appeared in Fritz \& Bruch (\cite{FB1998}).
Indeed, the scalegram is particularly conveniently implemented using the
wavelet transform results: In practice, it is the wavelet-summed
power for progressively increased time
scales as they change along the vertical axis
in Figs.~\ref{Morlet-6} or \ref{DOG-2}.
With $N$ points of input data,
for the scale (number of equidistant data intervals)
incremented as $S = 2, 4, 8, ...$,
the scalegram $V(S)$ for the scale $S$ is defined as:
$V(S) \propto (2S/N) \times \sum{W_i^2(S)}$, where $W_i(S)$ are the
individual wavelet transform components for a given scale.
If $V(S)$ shows a linear behaviour in logarithmic units
corresponding to $V(S) \propto S^{2\alpha}$,
then the underlying process is considered to be self-similar
with exponent $\alpha$.

For HD~37806, the DOG-2 wavelet transform
appears to provide the best description
of isolated and/or superimposed flares. In Fig.~\ref{DOG-2}
we directly see the intensifying wavelet strength
in the upward direction, towards longer scales.
When expressed as the scalegram in log-log units
(Figure~\ref{scale_disp}, the upper panel),
the progression takes the shape of a linear dependence
on time scales ranging approximately
between 0.25 days and 4 days. For the MOST-orbit averaged
data, the shorter time scales (shorter than the MOST orbit length)
are not accessible, while scales longer
than about 4 -- 5 days become affected by edge effects
of the full data sequence. Within the accessible range,
the logarithmic slope appears to be about 3
(not a formal fit, just a rough approximation), so that the noise process
is characterized by the exponent  $\alpha \simeq 1.5$.
The sign and the magnitude of the exponent mean that longer noise events
are stronger in intensity. Of course, any process with $\alpha > 0$
has such a property, but we note that this particular value of
the exponent is identical to that seen by Fritz \& Bruch (\cite{FB1998})
in many cataclysmic variables. However, for HD~37806, the time scales
involved are hours and days rather than minutes as observed in CV's.

\begin{figure}
\begin{center}
\includegraphics[width=80mm]{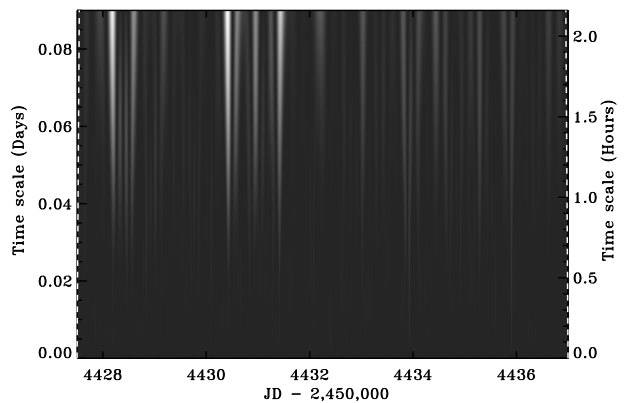}
\caption{The DOG-2 wavelet applied to the HD~37806 data sampled
at one minute intervals during the second half of the MOST
observations. Note the qualitatively different picture
from that in Figure~\ref{DOG-2}.
}
\label{DOG-2short}
\end{center}
\end{figure}

The MOST-orbit frequency sampling of the data at 0.07 d
sets an obvious limitation for the shortest time scales.
This could be partly remedied by using the
second half, 9.5 days long portion of the MOST observations
which were obtained at one minute intervals without repeated
scattered-light gaps. Still, even for this sequence,
some data dropouts were present
resulting in 97\% temporal coverage; those single,
occasional gaps were filled by spline interpolated points.
The DOG-2 wavelet transform looks very different without any
obvious indication of growing, coherent structures
(Figure~\ref{DOG-2short}). The small features which
we do see can be explained by minor instrumental discontinuities
and other data imperfections which became relatively more
important when the fluctuations in stellar flux were very small.

The wavelet transform and the related scalegram are excellent tools in
defining the {\it temporal scales\/} and
the noise-strength {\it progression with scale\/}.
But the {\it size or strength\/} of the fluctuations is difficult
to gauge. To relate the scalegram results to the observed
data scatter measured by the standard mean error $\sigma_{obs}$,
a simple analysis of the scatter was done for the whole
available MOST dataset: Values of the mean standard error
$\sigma_{obs}$ were calculated along the
time sequence for increased scales (intervals)
starting at single minutes and ending at 10 days, making sure that
each interval had a meaningful number of at least 5 points. Then, for
each scale $S$, the median $\sigma_{med}(S)$ value
was found from all individual values of $\sigma_{obs}$
as a characteristic noise number
for that scale. Although very simple, this approach
has led to surprisingly
well defined values of the typical fluctuations as
they varied with the length of the interval. The results are
plotted in the lower panel of Figure~\ref{scale_disp}
after being corrected for the observational random error $\sigma_R$ as
$\sigma^2_\star(S) = \sigma_{med}^2(S) - \sigma_R^2$.
The assumption for the MOST data was that $\sigma_R = 0.0028$ as
estimated for the shortest scales and in agreement with
the expected observational random error at $V \simeq 8$.
The results confirm the scalegram
progression and may even confirm that a change in the noise
characteristics was present around 0.25 days.

The same procedure of the $\sigma^2_\star$ determination
as for the MOST data was repeated for the
longer but more poorly sampled ESO data
(Section~\ref{eso} and Fig.~\ref{ESOlc}) and the ASAS data
(Section~\ref{asas} and Fig.~\ref{ASASlc}).
The MOST and the ground-based results
appear to agree very well and overlap in
Figure~\ref{scale_disp} for the stated
observational error of the ESO observations of $\sigma_R = 0.007$.
For the ASAS observations the error has been assumed to be
$\sigma_R = 0.017$, which is a less well defined number
but agrees with the typical accuracy of 0.015 to 0.02 mag.
The brightness fluctuations of HD~37806
appear to grow in size up to the accessible
scales of some two thousand days, confirming the ``red-noise''
characteristics of the variability.

\section{Discussion and interpretation}
\label{disc}

Because of the lack of spectroscopic support, our data cannot
unequivocally determine the cause of variability of HD~37806.
Some guidance can be found in extensive recent results for
T~Tauri stars to which Herbig Ae stars appear to be more
similar in terms of their activity than to Herbig Be stars
(Mottram et al.\ (\cite{mott2007}).
Of various processes identified in T~Tauri stars
(Bertout \cite{Bert1989}, Bertout et al.\ \cite{Bert1988},
Bouvier et al.\ \cite{Bouv2007}), one may consider:
(1)~photospheric spot rotational modulation,
(2)~variable obscuration by the dusty accretion disk,
(3)~instabilities in the accretion shock funnel on the stellar surface,
(4)~instabilities in the accretion disk. In addition,
Herbig Ae stars appear within the $\delta$~Sct
instability strip so that pulsations are also possible.

The wavelet analysis tells us that variability of HD~37806
takes the form of upward-directed spikes, rather than
dimming events. This suggests that it is caused
by accretion instabilities on the surface or in the space
adjacent to the star.
Rich and complex accretion phenomena in young stars have been
thoroughly described in the monograph of Hartmann (\cite{Hart2009}).
While considered by us as the most likely process from
the start, the accretion explanation has recently found
a very strong support in the
discovery of Wheelwright et al. (\cite{Wh2010}) that
the star is a binary system. This discovery
needs confirmation as the angular separation is
relatively small for the method used,
$0.025 \pm 0.001$ arcsec, and the data
for HD~37806 are of moderate quality. The implied
physical separation of components
for the distance of 300 -- 500~pc is of the order of
0.7 -- 1.0~AU indicating that
the accretion process involves relatively
large spatial dimensions; this would agree with our results
that the red-noise variability extends to temporal time
scales as long as several years (see below).

An explanation of the observed variability of HD~37806 by
stellar pulsations does not seem to be viable.
Light variations of HD~37806 in no way resemble stellar pulsations
observed for another Herbig Ae star HD~142666, which was also
observed by MOST (Zwintz et al. \cite{Zw2009}). Also, HD~37806
is hotter than the hot edge of the classical Delta Scuti
instability strip. Still, we carefully analyzed the MOST data
for HD~37806 in the frequency range where $\delta$~Scuti-like pulsations
would be expected, i.e.\ for frequencies between 5 and 80 c/d
or periods between about 5 hours and 20 minutes.
The only detectable, significant peaks in the amplitude spectrum
can be attributed to the instrumentally induced frequencies
which are well characterized for the MOST data.

The quasi-periodic signal with a period of about 1.5 d, which is
visible in the Fourier (Figure~\ref{MOSTfreq}) as well as
in the Morlet transform (Figure~\ref{Morlet-6}), finds the simplest
explanation in an accretion area periodically brought into view by
rotation of the star. Note, that this appears to be a bright,
weakly-defined signature rather than one corresponding
to a dark, sunspot-like feature.
The observed $V \sin i = 120 \pm 30$ km/s (Boehm \& Catala \cite{BC1995})
can be reproduced for this period -- assuming rigid rotation -- by
$R = (3.5 \pm 0.8\,R_\odot)/\sin i$.

The main finding of this work of self-similar brightening events
combining into chaotic red-noise
(Figures~\ref{DOG-2} and \ref{scale_disp})
may be interpreted as manifestations of instabilities in an accretion
disk around the star with progressively larger areas of the accretion disk
producing longer-lasting events.
The instabilities produce undetectable ($<$0.0003 of the mean flux)
variations for time scales of minutes, become very well
defined (at about 0.003 of the mean flux)
for time scales of hours, grow to $\simeq 0.010$ of
the mean flux at time scales of 1 --  10 days and continue increasing
for time scales of weeks to a few years where they reach 0.025
of the mean flux (see Figure~\ref{scale_disp}). Although
these variations are very well visible and characterized
thanks to the good quality of the observations,
they remain relatively small,
indicating that the star dominates in the mean brightness.

If interpreted by circumstellar events with time scales dynamically
related to sizes, for a $3\,M_\odot$ star, dimensions corresponding
to scales of hours to years would range between
a fraction of a stellar radius to $\simeq 3$~AU. Such a range of dimensions
may correspond to an accretion disk similar to those observed
in T~Tauri stars or one created by the matter falling from a close companion.
The interferometric observations of Eisner et al.\ (\cite{Eisn2004})
did not rule out the presence of a binary companion at
$< 1$~AU and were unable to determine the
inclination of the circumstellar material around the star.
The new spectro-astrometric method of Wheelwright et al.\ (\cite{Wh2010})
did detect a companion at $\le$~1~AU, but the detection is rather preliminary
and requires confirmation.

\begin{figure}
\begin{center}
\includegraphics[width=75mm]{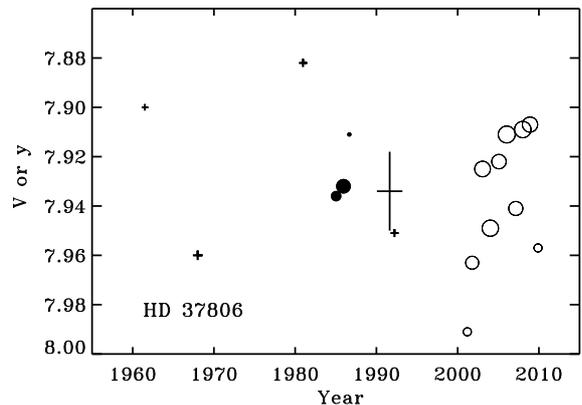}
\caption{$V$-band magnitudes of HD~37806 from the literature
(Hardie et al.\ \cite{Hard1964}, Warren \& Hesser \cite{WH1977},
de Winter et al.\ \cite{deW2001},
see Section~\ref{arch}; crosses), ESO (Manfroid et al.\ \cite{eso},
see Section~\ref{eso}; filled
circles) and ASAS (Section~\ref{asas}; open circles) with sizes of
symbols logarithmically scaled depending on the number of data points
in a given season (compare with Figures~\ref{ESOlc} and \ref{ASASlc}).
While the literature plotted data correspond
to typically 2 -- 5 observations obtained within a few days,
the ESO and ASAS observations correspond up to 80
observations per season. The large cross marks the mean Tycho-2 $V$
magnitude with its uncertainty
assigned here for the duration of the Hipparcos mission.
}
\label{long_term}
\end{center}
\end{figure}

Does the red-noise variability keep on growing at time scales
longer than 9 years covered by the ASAS (Section~\ref{asas})
project?
A comparison of $V$ magnitudes obtained from various sources
(Section~\ref{arch} and extending over about half a century,
is shown in Figure~\ref{long_term}. Although the data are of
a somewhat non-uniform quality, there is a clear indication
that variations of HD~37806 are
apparently small, not exceeding $\pm 0.05$ of the mean flux, which
is not much more than accuracy of some of the early observations.
The small spread in the measured $V$ magnitudes is in fact surprising
taking into account that some of the archival observations
were transformed from the $y$-band of the $ubvy$ photometry
and that the star is an emission-line object; both circumstances
might have easily increased the observational scatter through
problems of photometric transformations. We note that the
only indication of systematic changes is the presence of a
small brightness decrease which may have taken place sometime
between 1995 and 2000.

\section{Summary}
\label{summ}

Although the Herbig Ae/Be star
HD~37806 was a secondary object during the MOST observations
of the Sigma Ori field, the results may be of primary
importance to studies of variability of these young,
massive stars. Variability of HD~37806 appears to be
similar in general characteristics to that of
the T~Tauri star TW~Hya (Rucinski et al.\ (\cite{ruc2008})
in that both stars -- observed by MOST and analyzed in a
similar way -- showed brightness fluctuations apparently due to
instabilities in the matter accretion process. An accretion
disk may be fed by circumstellar matter or -- in the
case of HD~37806 -- most likely by a companion.
The process is unstable and
produces progressively larger variability amplitudes
at longer temporal scales,
a property which can be characterized as ``red noise''.

The variations in TW~Hya and HD~37806 differ in terms of
temporal as well as magnitude of variations.
For TW~Hya, the red noise contained discrete, relatively
large (typically 0.1 -- 0.2 of the mean flux), periodic
variations lasting typically a couple of weeks. They
evolved in their period-length towards shorter periods
(a well defined variation evolved from about 6 days to 3 days
in three weeks and another appeared at about 7 days).
For HD~37806, variations are much smaller
(typically 0.01 -- 0.02 of the mean flux) which may reflect a much
larger brightness of the star itself, and
we see no indications of periodic components similarly
drifting in period: A single, weak, periodic signature
of about 1.5 days did not show any temporal evolution in time
and may be related to inhomogeneities on this rapidly
rotating star. However -- importantly -- brightening events of
HD~37806 showed clear self-similar characteristics, growing in
intensity with time scale. The self-similar characteristics
of stochastic variability were established for
the time scale range of about 4 hours to 4 days.
The self-similarity exponent ($\alpha \simeq 1.5$) of the noise
was found to be very similar to that observed
in cataclysmic variables at much shorter
time scales of a minute.
The similarity to the CV's may be not accidental in view
of the recent result of Wheelwright et al. (\cite{Wh2010})
that HD~37806 is most likely a binary with the
component separation of $\le 1$~AU.

The archival data on the variability of HD~37806 suggest
its continuation -- with a possible moderation --
for long time scales.
The red noise characteristics do appear
in a simple analysis of the typical (median) standard
dispersion of the noise for progressively increased
time scales.
The large temporal and time-range extent of the combined
MOST, ESO and ASAS data permits one to see a well defined
progression in the amplitude -- scale
dependence up to intervals of a few years.
The red-noise fluctuations grow
for time scales ranging from minutes to a few years
from below 0.0003 to 0.025 of the mean flux.
The fragmentary and -- by their nature -- somewhat
non-uniform archival $V$ magnitude data do not suggest
any drastic noise increase ($<$~0.05) for still longer time scales
extending to half of a century.

\section*{Acknowledgments}
The Natural Sciences and Engineering Research Council of
Canada supports the research of DBG, JMM, AFJM, and SMR;
additional support for AFJM comes from FQRNT (Qu\'ebec).
RK and WWW are supported by the Austrian Space
Agency and the Austrian Science Fund.
KZ acknowledges support by the Austrian
{\it Fonds zur F\"orderung der wissenschaftlichen Forschung}
(FWF; project T335-N16) and is recipient of an APART
fellowship of the Austrian Academy of Sciences at the
Institute of Astronomy of the University of Vienna.
GP acknowledges support by the Polish MNiSW
grant N203 007 31/1328.

This research has made use of the SIMBAD database,
operated at CDS, Strasbourg, France and NASA's Astrophysics
Data System (ADS) Bibliographic Services.

We thank the reviewer for useful and constructive
comments and suggestions.


\begin{thebibliography}{}

\bibitem[1998]{vdA1998}
van den Ancker, M. E., de Winter, D., Tjin A Djie, H. R. E.,
1998, \aap, 330, 145

\bibitem[1995]{BC1995}
Boehm, T., Catala, C., 1995, \aap, 301, 155

\bibitem[1989]{Bert1989}
Bertout, C. 1989, Ann.\ Rev.\ Astr.\ Astroph., 27, 351

\bibitem[1988]{Bert1988}
Bertout, C., Basri, G., Bouvier, J. 1988, \apj, 330, 350

\bibitem[2007]{Bouv2007}
Bouvier, J., Alencar, S. H. P., Boutelier, T., et al.
2007, \aap, 463, 1017

\bibitem[1994]{brown1994}
Brown, A. G. A., de Geus, E. J., de Zeeuw, P. T.,
1994, \aap, 289, 101

\bibitem[2004]{Eisn2004}
Eisner, J. A., Lane, B. F., Hillenbrand, L. A.,  et al.,
2004, \apj, 613, 1049

\bibitem[1973]{feld1973}
Feldman, P. A., Purton, C. R., Marsh, K. A.,
1973, Nature Phys.\ Sci., 245, 7

\bibitem[1998]{FB1998}
Fritz, T., Bruch, A., 1998, \aap, 332, 586

\bibitem[1981]{guet1981}
Guetter, H. H., 1981, \aj, 86, 1057

\bibitem[1964]{Hard1964}
Hardie, R. H., Heiser, A. M., Tolbert, C. R. 1964, \apj, 140, 1472

\bibitem[2008]{haret2008}
Hareter, M., Reegen, P., Kuschnig, R., et al.
2008, CoAst, 156, 48

\bibitem[2009]{HK2009}
Harrington, D. M., Kuhn, J. R., 2009, \apjs, 180, 138

\bibitem[2009]{Hart2009}
Hartmann, L., 2009, Accretion Processes in Star Formation,
Cambridge University Press, 2nd Ed. (1st Ed, 1998)

\bibitem[2000]{Tycho2}                      
    H{\o}g, E., Fabricius, C., Makarov, V.V., Urban, S., Corbin, T.,
    Wycoff, G., Bastian, U., \\
    Schwekendiek, P., \& Wicenec, A.
    2000, \aap, 355L, 27

\bibitem[2007]{new-hip}
van Leeuwen, F., 2007, Hipparcos, the New Reductions of the
Raw Data; Astroph.\ Space Lib, Vol.350 (Springer)

\bibitem[1991]{eso}
Manfroid J., Sterken C., Bruch A., et al.
1991, A\&AS, 87, 481 (ESO Sci.\ Rep., 8)

\bibitem[2004]{M2004}
Matthews J.M., Kusching R., Guenther D.B., et al.
2004, Nature, 430, 51

\bibitem[1933]{mb1933}
Merrill, P. W., Burwell, C. G., 1933, \apj, 78, 87

\bibitem[1932]{mhb1932}
Merrill, P. W., Humason, K. L., Burwell, C. G., 1932, \apj, 76, 156

\bibitem[2007]{mott2007}
Mottram, J. C., Vink, J. S., Oudmaijer, R. D., Patel, M.,
2007, \mnras, 377, 1363

\bibitem[2006]{BP2006}
Paczy\'{n}ski, B., Szczygie{\l}, D., Pilecki, B., Pojma\'{n}ski, G.,
2006, \mnras, 368, 1311

\bibitem[1997]{perry1997}
Perryman, M. A. C. et al., 1997, \aap, 323, L49

\bibitem[1997]{Poj1997}
Pojma\'nski, G., 1997, Acta Astr., 47, 467

\bibitem[2002]{Poj2002}
Pojma\'nski, G., 2002, Acta Astr., 52, 397

\bibitem[2004]{Poj2004}
Pojma\'nski, G., 2004, Astron.\ Nachr., 325, 553

\bibitem[2005]{Poj2005}
Pojma\'nski, G., and Maciejewski, G.,  2005, Acta Astr., 55, 97

\bibitem[1978]{Press1978}
Press, W. H., 1978, Comm.\ Astrophys., 7, 103

\bibitem[2008]{ruc2008}
Rucinski, S. M., Matthews, J. M., Kuschnig, R., et al.
2008, MNRAS, 391, 1913

\bibitem[1993]{scarg1993}
Scargle, J.D., Steiman-Cameron, T., Young, K., et al.
Donoho, D.L., Crutchfield, J.P., Imamura, J.,
1993, \apj, 411, L91

\bibitem[2004]{sher2004}
Sherry, W. H., Walter, F. M., Wolk, S. J.,
2004, \apj, 128, 2316

\bibitem[2008]{sher2008}
Sherry, W. H., Walter, F. M., Wolk, S. J., Adams, N. R.,
2008, \apj, 135, 1616

\bibitem[1943]{SS1943}
Swings, P., Struve, O. 1943, \apj, 98, 91

\bibitem[1994]{The1994}
Th\'{e}, P.S., de Winter, D., P\'{e}rez, M.R.
1994, \aaps, 104, 315

\bibitem[1998]{TC1998}
Torrence, C., Compo, G. P.,
1998, Bull.\ Amer.\ Meteorological Soc., 79, 61

\bibitem[2005]{Vink2005}
Vink, J. S., Drew, J. E., Harries, T. J., Oudmaijer, R. D., Unruh, Y.
    2005, \mnras, 359, 1049

\bibitem[2003]{WM2003}
Walker G., Matthews J., Kuschnig R., et al.
2003, \pasp, 115, 1023

\bibitem[1977]{WH1977}
Warren, W. H., Hesser, J. E. 1977, \apjs, 34, 115

\bibitem[1998]{WW1998}
Waters, L. B. F. M., Waelkens, C.
1998, Ann.\ Rev.\ Astr.\ Astrophys., 36, 233

\bibitem[2010]{Wh2010}
Wheelwright, H. E., Oudmaijer, R. D., Goodwin, S. P.
2010, \mnras, 401, 1199

\bibitem[2001]{deW2001}
de Winter, D., van den Ancker, M.E., Maira, A., et al.
2001, \aap, 380, 609

\bibitem[2009]{Zw2009}
Zwintz, K., Kallinger, T., Guenther, D. B. et al.
2009, \aap, 494, 1031

\end{thebibliography}
\end{document}